\documentclass[aps,pra,twocolumn,superscriptaddress]{revtex4-1}
\usepackage{graphicx}% Include figure files
\usepackage{dcolumn}% Align table columns on decimal point
\usepackage{bm}% bold math
\usepackage{amsmath}
\usepackage[english]{babel}
\begin{document}

\title{Gate-controlled quantum dots and superconductivity in planar germanium}
\author{N.W. Hendrickx}
\email{n.w.hendrickx@tudelft.nl}
\author{D.P. Franke}
\affiliation{QuTech and Kavli Institute of Nanoscience, Delft University of Technology, PO Box 5046, 2600 GA Delft, The Netherlands}
\author{A. Sammak}
\affiliation{QuTech and the Netherlands Organisation for Applied Scientific Research (TNO), Stieltjesweg 1, 2628 CK Delft, The Netherlands}
\author{M. Kouwenhoven}
\author{D. Sabbagh}
\author{L. Yeoh}
\affiliation{QuTech and Kavli Institute of Nanoscience, Delft University of Technology, PO Box 5046, 2600 GA Delft, The Netherlands}
\author{R. Li}
\author{M.L.V. Tagliaferri}
\affiliation{QuTech and Kavli Institute of Nanoscience, Delft University of Technology, PO Box 5046, 2600 GA Delft, The Netherlands}
\author{M. Virgilio}
\affiliation{Dipartimento di Fisica ``E. Fermi'', Universit\`a di Pisa, Largo Pontecorvo 3, 56127 Pisa, Italy}
\author{G. Capellini}
\affiliation{Dipartimento di Scienze, Universit\`a degli studi Roma Tre, Viale Marconi 446, 00146 Roma, Italy}
\affiliation{IHP, Im Technologiepark 25, 15236 Frankfurt (Oder), Germany}
\author{G. Scappucci}
%\email{g.scappucci@tudelft.nl for material requests}
\author{M. Veldhorst}
\email{m.veldhorst@tudelft.nl}
\affiliation{QuTech and Kavli Institute of Nanoscience, Delft University of Technology, PO Box 5046, 2600 GA Delft, The Netherlands}
\date{\today}

\maketitle
\setlength{\textwidth}{183mm}

\textbf{Superconductors and semiconductors are crucial platforms in the field of quantum computing \cite{nakamura_coherent_1999,petta_coherent_2005}. They can be combined to hybrids, bringing together physical properties \cite{basov_towards_2017} that enable the discovery of new emergent phenomena \cite{mourik_signatures_2012} and provide novel strategies for quantum control \cite{larsen_semiconductor-nanowire-based_2015}. The involved semiconductor materials, however, suffer from disorder, hyperfine interactions or lack of planar technology. Here we realise an approach that overcomes these issues altogether and integrate gate-defined quantum dots and superconductivity into a material system with strong spin-orbit coupling. In our germanium heterostructures, heavy holes with mobilities exceeding 500,000 cm$^2$/Vs are confined in shallow quantum wells that are directly contacted by annealed aluminium leads. We demonstrate gate-tunable superconductivity and find a characteristic voltage $I_cR_n$ that exceeds 10 $\mu$V. Germanium therefore has great promise for fast and coherent quantum hardware and, being compatible with standard manufacturing, could become a leading material in the quantum revolution.}

The group-IV semiconductors Si and Ge come with central advantages for the realisation of spin quantum bits (qubits). Not only has their purity and technology been refined to a formidable level, they also possess an abundant isotope with zero nuclear spin \cite{itoh_high_1993, itoh_isotope_2014}, enabling spin qubits to reach extremely long coherence times \cite{veldhorst_addressable_2014, sigillito_electron_2015} and high fidelity \cite{yoneda_quantum-dot_2017}. These powerful properties have led to demonstrations of two-qubit logic gates \cite{veldhorst_two-qubit_2015, zajac_resonantly_2018} and quantum algorithms \cite{watson_programmable_2017}. The exchange interaction that is central in these demonstrations is local and cannot directly be used to couple qubits at a distance. Instead, long-range coupling of spin qubits is being explored by incorporating superconductivity and in a first step strong spin-photon coupling has been achieved \cite{mi_strong_2017, samkharadze_strong_2018}. 

Hole quantum dots in Ge are particularly promising in this context. Ge has the highest hole mobility of all known semiconductors \cite{pillarisetty_academic_2011} and strong spin-orbit coupling facilitates electrical driving for fast qubit operations. Experiments have shown readout of holes in Ge/Si nanowires \cite{lu_one-dimensional_2005, hu_ge/si_2007} and self-assembled quantum dots \cite{ares_nature_2013}, and promising spin lifetimes have been found \cite{hu_hole_2012}. In addition, the strong Fermi level pinning at the valence band edge leads to ohmic behaviour for all \mbox{metal - ($p$-type) Ge contacts} \cite{dimoulas_fermi-level_2006}. 
The resulting strong coupling between metal and semiconductor enables the fabrication of hybrid devices of quantum dot and superconducting structures \cite{katsaros_hybrid_2010,larsen_semiconductor-nanowire-based_2015}. 

Now the crucial next step is the construction of a planar platform, compatible with the silicon fabrication technology, that can bring together strong spin-orbit coupling, low disorder, and avenue for scaling. Here, we address this challenge and present the formation of a quantum dot in a planar Ge quantum well. Further, we implement direct Al-based ohmic contacts that eliminate the need for dopant implantation. In addition, the Al leads can proximity-induce superconductivity in the quantum well and we can control the associated supercurrent by tuning the electrical gates.

\begin{figure*}%
	\includegraphics[width =183mm]{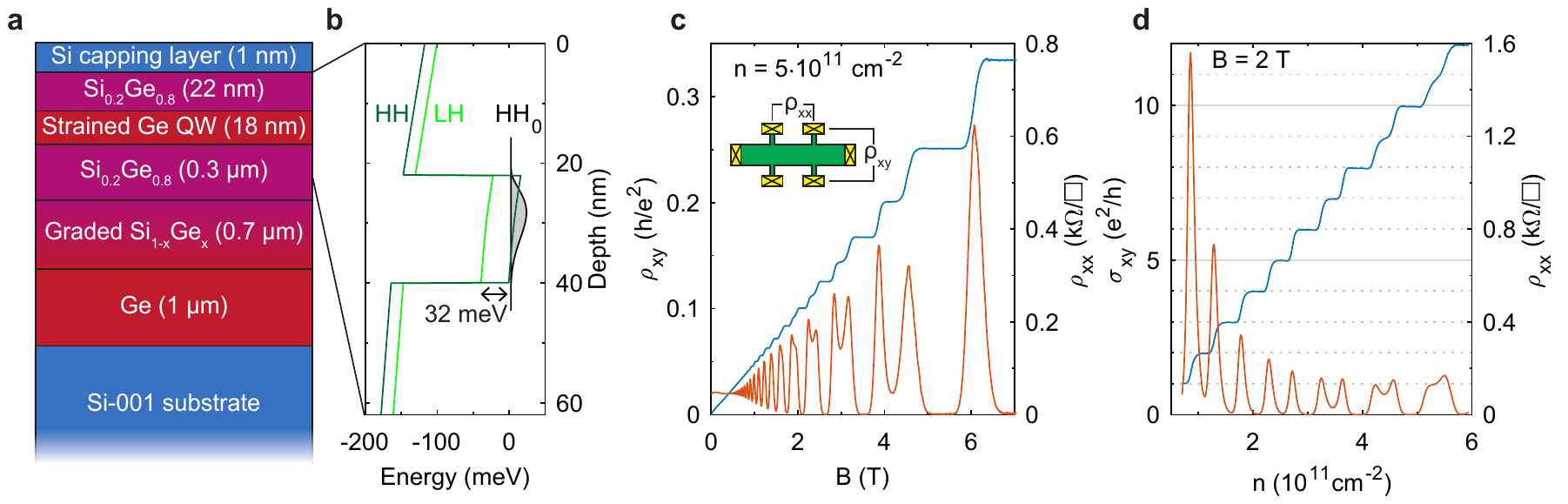}%
	\caption{\textbf{Ge/SiGe heterostructure and magnetotransport measurements.} 
	\textbf{a}, Schematic representation of the full heterostructure (dimensions not to scale). \textbf{b}, Simulated band structure within the SiGe-Ge-SiGe heterostructure, showing the strong confinement of heavy holes in the Ge. 
	\textbf{c},\textbf{d}, Magnetotransport measurements in a perpendicular magnetic field $B$, showing clear Shubnikov-de Haas oscillations and quantized Hall conductance as a function of $B$ (\textbf{c}) and the carrier sheet density $n$ (\textbf{d}) as converted from the top gate potential using low-field Hall effect data.}
	\label{fig:QW}%
\end{figure*}

Si and Ge are completely miscible and the lattice constant of their alloy, SiGe, varies continuously between its constituents. This is exploited by using strain-relaxed compositionally graded SiGe layers as virtual substrates to define Ge/SiGe heterostructures as shown schematically in \mbox{Fig.~\ref{fig:QW}a}. We use a high-throughput reduced-pressure chemical vapour deposition (RP-CVD) reactor to grow the complete heterostructure in one deposition cycle on a 100 mm Si(001) substrate. The Ge quantum well is deposited pseudomorphically on a strain-relaxed Si$_{0.2}$Ge$_{0.8}$/Ge/Si virtual substrate obtained by reverse grading \cite{shah_reverse_2010} (see methods). The resulting in-plane compressive strain in the quantum well splits the valence band states, which further increases the hole mobility \cite{schaffler_high-mobility_1997}. The quantum well layer is separated from the surface by a Si$_{0.2}$Ge$_{0.8}$ spacer and a Si cap.

Because of the type-I band alignment in Ge-rich Ge/SiGe heterostructures, the valence band maximum is energetically higher in Ge than in SiGe, such that holes accumulate in the Ge quantum well \cite{virgilio_type-i_2006}. Heavy and light hole electronic states (HH and LH, resp.) and the related band profiles have been calculated by solving the Schr\"odinger-Poisson equation for low temperatures as a function of vertical electric field. 
In our simulations both a multi valley effective mass approach \cite{virgilio_physical_2014} and an atomistic tight-binding model \cite{busby_near-_2010} have been used, obtaining consistent data. This agreement supports the accuracy of our numerical results. The calculated valence band edge profile is shown in Fig.~\ref{fig:QW}b, where the wavefunction of the fundamental state HH$_0$, which has heavy hole symmetry and is well confined in the Ge layer, is also sketched. The strain induced splitting of the HH and LH band edges in the Ge region increases the energy of the fundamental light hole state LH$_0$ (not shown) which is found at about 46 meV above HH$_0$. Furthermore, we estimate a value of 14 meV for the first excited heavy hole level HH$_1$ (not shown). Notice that these energy splittings are significantly larger than those resulting from the valley interaction in Si/SiO$_2$ and Si/SiGe devices and thus excellent conditions for the operation of spin qubits are expected in this material.

The hole mobility is measured in magnetotransport experiments at 50 mK using a heterostructure field effect transistor as shown in the inset of Fig.~\ref{fig:QW}c. Here, the yellow boxes indicate metallic ohmic contacts to the quantum well created by diffusion into the top SiGe layer and the green structure represents an isolated Hall-bar-shaped top gate which is used to control the hole density in the quantum well. 
Clear Shubnikov-de Haas oscillations with zero-resistivity minima and Zeeman splitting of the Landau levels at higher fields are observed in the longitudinal resistance $\rho_{xx}$ as a function of the magnetic field (Fig.~\ref{fig:QW}c). 
Furthermore, we observe flat quantum Hall effect plateaus at values $1/\nu$ for integer $\nu$ in the transverse resistance $\rho_{xy}$ in units of $h/e^2$, where $h$ is Planck's constant and $e$ the elementary charge. 
The investigated heterostructures support a density of up to $6\times 10^{11}$ cm$^{-2}$ and have a maximum mobility of more than $500,000~ \mathrm{cm}^2/\mathrm{Vs}$, providing new benchmarks for SiGe structures. Figure \ref{fig:QW}e shows the Hall conductivity $\sigma_{xy}$ in units of $e^2/h$ as a function of the hole density $n$, controlled by the top gate potential. 
Again, zero-resistivity minima in $\rho_{xx}$ and clear linear quantization steps in $\sigma_{xy}$ are observed. This demonstrates the control of the hole density over a large range using the top gate, which is a central prerequisite for the definition of electrostatically defined quantum dots.

\begin{figure*}%
	\includegraphics[width=183mm]{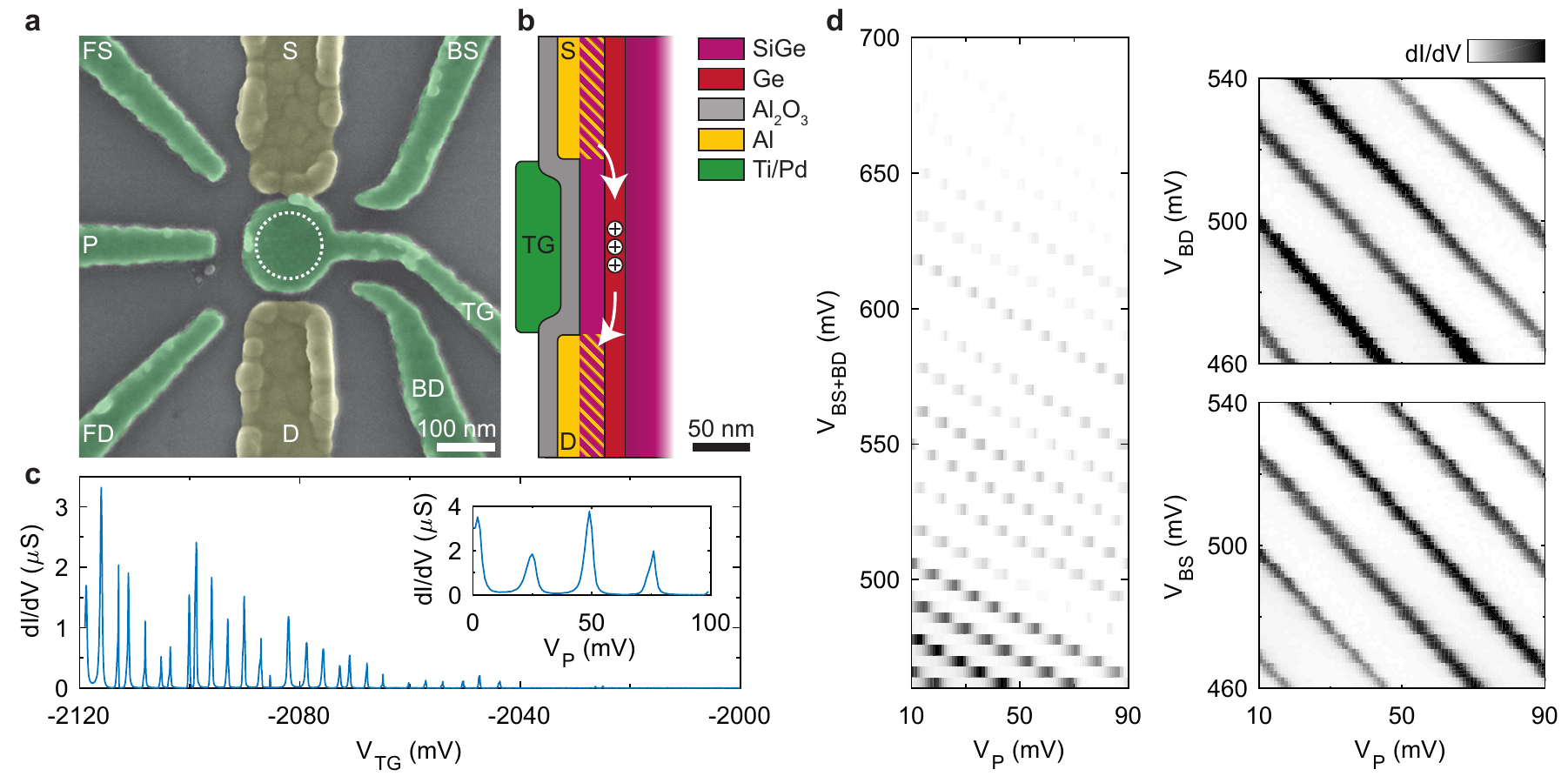}%
	\caption{\textbf{Fabrication and operation of a gate-defined quantum dot.} 
	\textbf{a}, False-coloured SEM image of the quantum dot device. The quantum dot is defined under the top gate TG (dotted circle) and its occupancy can be controlled by the central plunger gate P. 
	BS and BD correspond to source and drain barriers, respectively, FS and FD are finger gates for additional control.
	\textbf{b}, Schematic of the device gate layers, showing the top gate and the ohmic contacts achieved by in-diffusion of Al. 
	\textbf{c}, Transport measurements showing Coulomb oscillations as a function of the top gate and the plunger gate (inset). 
	\textbf{d}, Influence of the barrier gates BS and BD on the observed current, demonstrating the tunnelling rate control using these gates. The coupling of the two individual gates to the quantum dot is nearly identical, emphasising the excellent homogeneity reached in this system.}%
	\label{fig:nano}%
\end{figure*}

A scanning electron microscope image of the quantum dot nano structure is shown in Fig.~\ref{fig:nano}a. Here, the ohmic Al contacts are coloured in yellow and the isolated Ti/Pd gates are shown in green.
In a first step, the Al contacts to the Ge quantum well are defined by electron beam lithography and local etching of the Si capping layer and thermal evaporation of Al. Subsequently, a Al$_2$O$_3$ gate dielectric is grown by atomic layer deposition at 300 $^\circ$C, which also serves as an annealing step to enable the diffusion of Al into the SiGe spacer.
In the Pd gate layer, we design a circular top gate between the two Al leads under which a single quantum dot will be formed. In addition, a central plunger gate P is included to control the dot occupation, as well as barrier gates (BS and BD) and additional finger gates (FS and FD) in all four corners of the device. These allow for additional control of the dot size and the tunnelling rates between the quantum dot and the source and drain leads.

A conceptual drawing of the device cross section is shown in Fig.~\ref{fig:nano}b, where the diffused Al leads that contact the Ge quantum well are indicated by the stripe-pattern regions. Because of the ohmic nature of the contact, the transport through the quantum dot can be measured without the need for additional reservoir gates and dopant implantation. As a result, no annealing step at temperatures higher than the quantum well deposition temperature ($500^\circ$ C) is needed during the fabrication, avoiding harmful Ge/Si intermixing at the interface \cite{dobbie_ultra-high_2012}.

When measuring the source drain conductance $dI/dV$ as a function of the top gate voltage, conductance peaks are expected when the dot potential $\mu$ is aligned between the source and drain potentials such that holes are transported through the dot, which are the so-called Coulomb oscillations (cf.~Fig.~\ref{fig:nano}c).
This is shown in Fig.~\ref{fig:nano}d, where $dI/dV$ was measured as a function of the top gate voltage $V_\mathrm{TG}$. The expected oscillations are observed, which is a clear sign of the formation of a quantum dot. 
From the period of these oscillations we can extract a top gate capacitance of $\sim 56$ aF, which is in very good agreement with the expected capacitance of 52 aF of a parallel plate capacitor using the lithographic dimensions of the top gate. 
When TG is tuned to the quantum dot regime, similar oscillations are observed as a function of the plunger gate voltage $V_\mathrm{P}$. 
Here, a larger spacing of the Coulomb peaks is observed, corresponding to a gate capacitance of $\sim 6.4$ aF, in agreement with the expected weaker coupling of P to the quantum dot. Note that because of the structure of the source and drain leads no additional tuning of the device is necessary. Equivalent to a classical transistor, the quantum dot can be defined using a single gate (TG), which bodes well for the scaling up of qubits in this system.

Using the two barrier gates BS and BD, the tunnelling rates between the dot and the source and drain reservoirs can be manipulated. 
This is demonstrated in Fig.~\ref{fig:nano}e, where the peak conductance lines in $dI/dV$ fade out as a function of the positive barrier voltages. 
When BS and BD are measured separately, it becomes clear that the coupling of each of the two barriers and the plunger to the quantum dot is nearly identical. 
This confirms that the quantum dot is formed in a central position under TG and that, as an effect of low disorder in the heterostructure, a very high level of symmetry and control is achieved.

\begin{figure*}%
	\includegraphics[width=183mm]{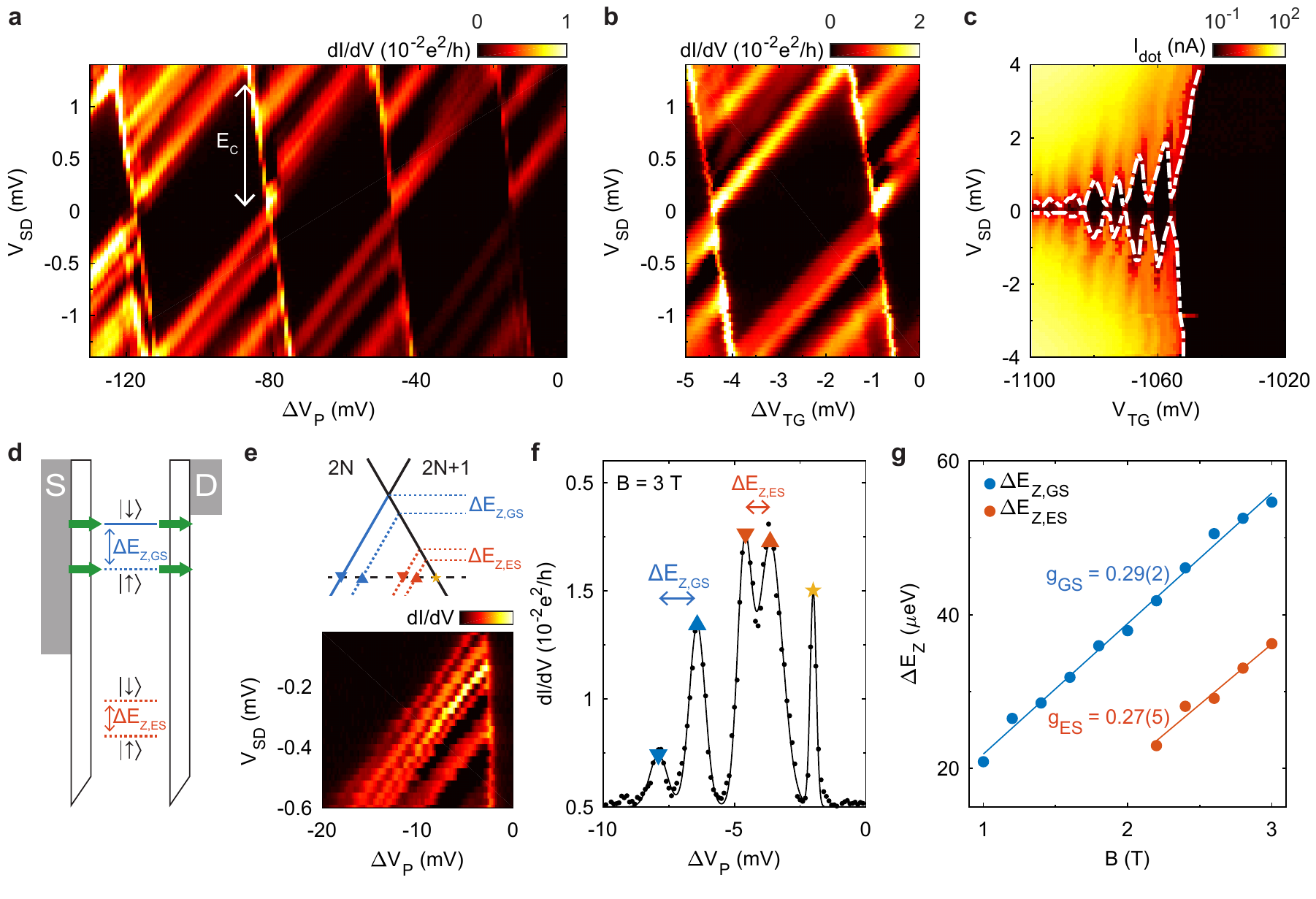}%
	\caption{\textbf{Bias spectroscopy and g-factor measurements.} \textbf{a}, \textbf{b}, Colour plots of bias spectroscopy as a function of $V_\mathrm{P}$ (\textbf{a}) and $V_\mathrm{TG}$ (\textbf{b}), showing Coulomb diamonds with $E_C = 1.3(1)$ meV. 
		\textbf{c}, Bias spectroscopy of a second device (see main text), demonstrating a quantum dot in the few hole regime. Dashed contour lines correspond to $I_\mathrm{dot}= 0.2$ nA.
		\textbf{d}, Schematic drawing of the Zeeman splitting $\Delta E_\mathrm{Z}=g\mu_\mathrm{B}B$ of the quantum dot levels, illustrating the hole transport via the Zeeman-split energy levels. 
		\textbf{e}, Bias spectroscopy showing the line splitting in a 3 T in-plane field. 
		\textbf{f}, Differential conductance as a function of $\Delta V_{\mathrm{P}}$ at $V_{SD}=-0.3$ mV. Solid line corresponds to a fit to the data using the sum of five Gaussian profiles.
		\textbf{g}, Energy splittings for the ground and excited state as a function of $B$. Solid lines are linear fits to the data yielding the corresponding $g$-factors.}
\label{fig:diamonds}
\end{figure*}
To further characterise the quantum dot, we measure $dI/dV$ as a function of $V_\mathrm{P}$ and the DC bias voltage $V_\mathrm{SD}$. As shown in Fig.~\ref{fig:diamonds}a, Coulomb diamonds are observed \cite{hanson_spins_2007}.
From the height and width of these diamonds, the charging energy $E_C$ and the lever arm of the corresponding gate $\alpha$ can be extracted. In the regime shown here ($V_\mathrm{BS}=V_\mathrm{BD}=550~\text{mV}$ and $V_\mathrm{FD}=600~\text{mV}$), we find $\alpha_\mathrm{P}= 0.037(3)$ and $E_C=1.3(1)$ meV.

Similar diamonds are observed as a function of $V_\mathrm{TG}$. Here, the lever arm is found to be significantly larger ($\alpha_\mathrm{TG} =0.41(3)$), as is expected because the dot is formed directly under TG.
A substructure is clearly visible in the conducting areas. These additional lines could correspond to either charge transport via excited states of the dot or to a modulation of the density of states within the source and drain reservoirs \cite{escott_resonant_2010}.

We further measure a second device where an additional top gate TG2 is implemented, which is set to high negative voltage ($V_\mathrm{TG2}=-2500$ mV) and acts as an extended source reservoir in the operation mode applied here. 
Similar to the first device, a dot is formed under the other top gate. When the barrier and finger gates are set to rather high positive voltages ($V\approx 1000$ mV), this dot can be measured in the few hole regime, as demonstrated in Fig.~\ref{fig:diamonds}c, where a clear increase of $E_C$ with lower occupation is observed. 
%When TG is raised above $-1050$ mV, the Coulomb diamond pattern abruptly ends, suggesting that we have fully depleted the dot.

When an external magnetic field $B$ is applied, the energy levels of spin degenerate states are expected to split as a result of the Zeeman effect \cite{hanson_spins_2007} (cf.~Fig.~\ref{fig:diamonds}d). 
This becomes apparent as a splitting of the conductance lines related to odd hole occupations $2N+1$ of the quantum dot, as shown in Fig.~\ref{fig:diamonds}e for the first device in an in-plane magnetic field $B = 3$ T.
Both the ground state and the excited state are subject to this splitting, which we extract by fitting the observed conductance for \mbox{$V_{\text{SD}}=-0.3$ mV} using Gaussian profiles and multiplying the splitting in voltage with the measured lever arm $\alpha_\mathrm{P}$ (Fig.~\ref{fig:diamonds}f).

A linear trend is observed as a function of the applied magnetic field as shown in Fig.~\ref{fig:diamonds}g. Note that small splittings $\Delta E_Z< 20~\mu$eV could not be resolved because of the finite width of the conductance peaks. 
We find the effective $g$-factors $g_\mathrm{GS}= 0.29(2)$ and $g_\mathrm{ES}=0.27(5)$ for the ground state and excited state, respectively. For the excited state, our data point to either a strong non-linearity at lower fields or a significant zero-field spin splitting $\Delta E_{\mathrm{Z}0}\approx -11~\mu eV$.
The $g$-factor of the pure heavy hole state is expected to vanish completely for an in-plane field. However, the additional confinement of the holes in the $x$,$y$-plane leads to a significant admixture of light hole states and a non-zero in-plane Zeeman splitting \cite{nenashev_wave_2003}.
%We further cannot exclude a slight misorientation of our device, which could have a notable influence on the observed $g$-factor because of the expected large out-of-plane Zeeman effect.

The observed spin splitting identifies this substructure as belonging to the first excited state rather than being connected to the reservoir. The measured energy splitting with respect to the ground state $\Delta E\approx 100~\mu$eV remains unchanged as a function of magnetic field strength.
It can be compared to the expected level splitting for a two-dimensional quantum dot with area $A$ and effective hole mass $m^*$, which is given by $\Delta E=\pi\hbar^2/m^*A$ \cite{kouwenhoven_electron_1997}. 
From the temperature dependence of the Shubnikov-de Haas oscillations measured for the Hall-bar structures we find $m^*=0.08m_e$, with the electron mass $m_e$, and with our device geometry ($A=0.019~\mu\mathrm{m}^2$) we obtain $\Delta E \approx 150~\mu$eV, in good agreement with the measured value.

\begin{figure}%
	\includegraphics[width=1\columnwidth]{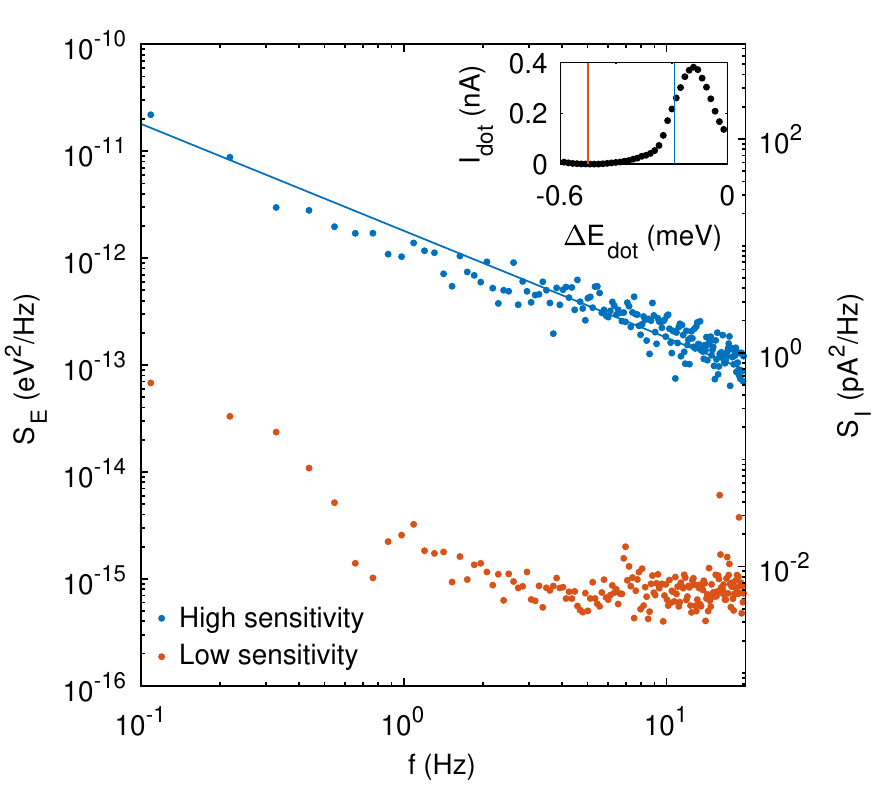}%
	\caption{\textbf{Quantum dot charge noise spectrum.}
	Current noise power spectrum for the quantum dot in transport (blue dots) and in Coulomb blockade (orange dots) as indicated in the inset. Solid line corresponds to an apparent linear fit to the data, yielding a slope of -1.00(4).}%
	\label{fig:chargenoise}%
\end{figure}

The creation of ohmic leads by the diffusion of Al in the direct vicinity of the quantum dot could be suspected of creating additional charge traps that have a negative influence on the coherence of a potential qubit. 
To quantify this effect, we measure the charge noise acting on our quantum dot by measuring the transport current $I_\mathrm{dot}$ in a sensitive region, i.e.~on the slope of a Coulomb peak. A 100-s-long time trace of $I_\mathrm{dot}$ is acquired at a sampling rate of 30.5 kHz and is decomposed into 10 traces of equal lengths. The discrete Fourier spectra obtained from these traces are averaged, yielding the noise spectral density $S_E$ presented in Fig.~\ref{fig:chargenoise} in comparison to the spectrum measured in a low sensitivity region (Coulomb blockade). The difference between the two traces confirms that the measured low frequency noise spectrum is indeed dominated by charge noise acting on the quantum dot. The noise spectrum follows a $1/f$ trend for low frequencies (solid line in Fig.~\ref{fig:chargenoise}), consistent with results on similar structures \cite{Basset_Evaluating_2014, Freeman_Comparison_2016}. We find the equivalent detuning noise at 1 Hz to be $1.4~\mu$eV/$\sqrt{\text{Hz}}$.
This compares well to noise figures at 1 Hz in other materials, such as 7.5, 0.5, or \mbox{2.0 $\mu$eV/$\sqrt{\mathrm{Hz}}$} in GaAs \cite{Basset_Evaluating_2014}, Si/SiO$_2$ \cite{Freeman_Comparison_2016} and Si/SiGe \cite{Freeman_Comparison_2016}, respectively, showing the suitability of our approach for the creation of low-noise qubits.

\begin{figure}%
	\includegraphics[width=1\columnwidth]{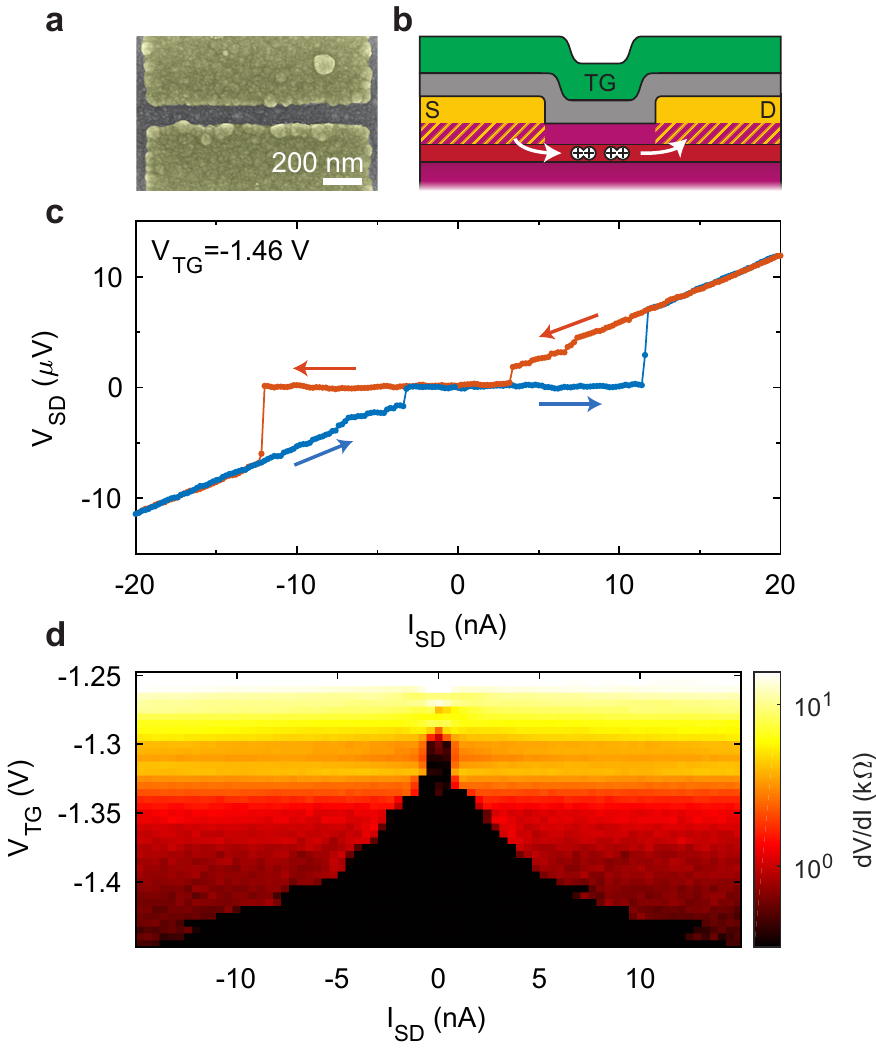}%
	\caption{\textbf{Tunable induced superconductivity in the Ge quantum well.}
    \textbf{a}, SEM image of the JoFET device with a gap of 100 nm. \textbf{b}, Device schematic using the colour scheme of Fig.~\ref{fig:nano}b. Superconducting transport through the quantum well is illustrated.
	\textbf{c}, Source-drain voltage as a function of the applied current at a top gate voltage of $V_{\text{TG}}=-1.46$ V. A flat plateau with zero resistance is observed up to a critical current of $I_c\approx 12$ nA as a clear signature of a supercurrent through the device.
    \textbf{d}, The critical current can be modulated by tuning the top gate voltage, as shown in the colour plot of the current dependence of the device resistance as a function of $V_\text{TG}$.}
	\label{fig:super}%
\end{figure}

The direct ohmic contact of the Al leads to the Ge quantum well can proximity-induce superconductivity in the semiconductor system. To demonstrate this effect, we fabricate a Josephson Field Effect Transistor (JoFET) device where two superconducting leads with a separation of 100 nm are overlaid with a single top gate. A SEM image of the device is shown together with a schematic drawing of the layer structure in Fig.~\ref{fig:super}a,b. When the source-drain voltage is measured as a function of the sourced current $I_\mathrm{SD}$ at negative top gate potential, a zero-resistance plateau is observed as a clear sign of a supercurrent. This is shown in Fig.~\ref{fig:super}c, where the blue and orange traces represent measurements in different sweep directions, respectively. Sweeping towards higher absolute values of $I_\mathrm{SD}$, the critical current $I_c$ is reached at $\sim 12$ nA. Beyond these values, linear (ohmic) behaviour is observed. A hysteresis can be observed when sweeping back, which can be expected for proximity induced SNS-junctions.
To demonstrate that this supercurrent is induced in the quantum well, we measure the observed bias current dependence of the source-drain resistance $dV/dI$ as a function of the top gate potential $V_\mathrm{TG}$. To avoid the influence of the hysteresis, two separate sweeps towards positive and negative currents are measured, the combined data is shown in Fig.~\ref{fig:super}d. The critical current, which defines the borders of the zero resistance range observed as black region in the colour plot, is reduced towards increasing $V_\mathrm{TG}$. Above \mbox{$V_\mathrm{TG}\approx -1.27$ V} the supercurrent vanishes completely. For decreasing gate potential, we find that the increases in $I_c$ is more significant than the reduction in the normal resistance $R_n$. The characteristic voltage $I_cR_n$ hence increases and reaches values higher than $\sim 10~\mu$V.

In conclusion, we have shown the operation of a hole quantum dot in a planar low-disorder buried Ge quantum well with a record hole mobility. The Al ohmic leads to the quantum dot significantly simplify the fabrication and tuning processes without an increase of the measured charge noise in comparison to other systems. The strong capacitive coupling between the superconductor and the quantum dot makes this system ideal for reaching strong spin-photon coupling, while the strong spin-orbit coupling present in heavy holes could be exploited for fast electrical qubit driving. Furthermore, the demonstration of gate-tunable superconductivity opens up new research directions including Majorana modes \cite{mourik_signatures_2012} and gatemons \cite{larsen_semiconductor-nanowire-based_2015}. Hole quantum dots in planar Ge constitute thereby a versatile platform that can leverage semiconductor manufacturing to advance and broaden the field of quantum computing. 

\section*{Acknowledgements}
The authors acknowledge support through a FOM Projectruimte of the Foundation for Fundamental Research on Matter (FOM), associated with the Netherlands Organisation for Scientific Research (NWO).

\section*{Methods}
\small The Ge/SiGe heterostructures were grown in one deposition cycle in an Epsilon 2000 (ASMI) RP-CVD reactor on a 100 mm $n$-type Si(001) substrate (resistivity 5 $\Omega$cm). The growth starts with a 1-$\mu$m-thick layer of Ge, using a dual-step process with initial low-temperature (400$^\circ$ C) growth of a Ge seed layer followed by a higher temperature (625$^\circ$ C) overgrowth of a thick relaxed Ge buffer layer. A cycle anneal at 800$^\circ$ C is performed to promote full relaxation of the Ge. Subsequently, a 700 nm reverse-graded Si$_{1-x}$Ge$_x$ layer is grown at 800$^\circ$ C with $x$ changing linearly from 1 to 0.8. A relaxed 300 nm buffer Si$_{0.2}$Ge$_{0.8}$ is then grown in two steps at an initial high temperature of 800$^\circ$ C followed by a low temperature growth at 500$^\circ$ C. The growth continues with a 16-nm-thick strained Ge quantum well, a 22-nm-thick Si$_{0.2}$Ge$_{0.8}$ barrier, and a 1-nm-thick Si cap. The final layers are all deposited at a temperature of 500$^\circ$ C.

The photolithography process for fabricating the Hall bar field effect transistors includes the following steps: mesa etching, deposition of 60-nm-thick Pt layer ohmic contacts, and atomic layer deposition (ALD) at 300$^\circ$ C of a 30-nm-thick Al$_2$O$_3$ dielectric to isolate a Ti/Au top gate (thicknesses 10/150nm).

For the quantum dot, the contact and gate structures are created by electron beam lithography, electron beam evaporation of Al and Ti/Pd and lift-off.
Following the Al contact layer (20 nm), ALD is used to grow $17$ nm of Al$_2$O$_3$ at $300~^\circ$C as a gate dielectric, followed by the Ti/Pd (5/35 nm) gate structures.
For the JoFET device, the same process is followed with layer thicknesses 30 nm (Al), 25 nm (Al$_2$O$_3$) and 5/35 nm (Ti/Pd).

Magnetotransport data has been obtained in a $^3$He dilution refrigerator with a base temperature of $50$ mK, equipped with a $9$ T magnet.
All quantum dot and JoFET measurements were performed in a $^3$He dilution refrigerator with a base temperature of $<10$ mK, equipped with a $3$ T magnet. Quantum dot confinement has been observed in two different devices and during three thermal cycles. All quantum dot measurements of $dI/dV$ are performed using lock-in amplification with typical modulation amplitude and frequency of $\delta V_\mathrm{SD}=10-100~\mu$V and $f_\mathrm{mod}=73.5$ Hz, respectively.
The JoFET devices were measured in a four-point configuration, sourcing a current and measuring the potential across the superconducting junctions. The plotted voltage is corrected by a small offset of the measurement electronics. The differential resistance is measured using lock-in amplification with typical a modulation amplitude of $\delta I_\mathrm{SD}=0.3$ nA. \linebreak

%\section*{Author contributions}
%{\small N.W.H.~fabricated the quantum dot devices and N.W.H.~and D.P.F.~performed the experiments. A.S.~prepared the heterostructures, D.S.~fabricated the Hall bar transistors measured at low temperature by D.S., L.Y., and G.S. N.W.H., D.P.F., G.S.~and M.V. analysed the data, M.Vi.~and G.C.~carried out additional analysis. N.W.H.~and D.P.F.~wrote the manuscript with input from all authors. M.V.~and G.S.~conceived and supervised the project.}

\bibliography{germanium}

\end{document}